\documentclass[letterpaper,11pt]{article}
\usepackage{jheppub}
\usepackage{epstopdf}
\usepackage{slashed}
\usepackage{graphicx} 
\usepackage{soul}
\usepackage{amsmath}
\usepackage{multirow}
\usepackage{tablefootnote}
\usepackage{comment}
\usepackage{color}
\usepackage[utf8]{inputenc}
\usepackage{caption,subcaption}
\usepackage[normalem]{ulem}
\usepackage[T1]{fontenc}

\usepackage{graphicx}
\usepackage{graphbox}
\usepackage{amsmath}
\usepackage{amsfonts}
\usepackage{amssymb}
\usepackage{cleveref}
\usepackage{color}
\usepackage{xcolor}

\newcommand{\beq}{\begin{equation}}
\newcommand{\eeq}{\end{equation}}
\newcommand{\bea}{\begin{eqnarray}}
\newcommand{\eea}{\end{eqnarray}}
\newcommand{\ba}{\begin{array}}
\newcommand{\ea}{\end{array}}

\def\m1{M_1}
\def\m2{M_2}
\def\m3{M_3}

\def\ch10{\tilde \chi^0_1}

\def\to{\rightarrow}

\newcommand{\lsim}{\mathrel{\mathop{\kern 0pt \rlap
  {\raise.2ex\hbox{$<$}}}
  \lower.9ex\hbox{\kern-.190em $\sim$}}}
\newcommand{\gsim}{\mathrel{\mathop{\kern 0pt \rlap
  {\raise.2ex\hbox{$>$}}}
  \lower.9ex\hbox{\kern-.190em $\sim$}}}
\preprint{PITT-PACC-2201}
\title{\Large{\bf WIMP Dark Matter at High Energy Muon Colliders\\
$-$A White Paper for Snowmass 2021} 
}

\author[a]{Tao Han,}
\author[b]{Zhen Liu,}
\author[c]{Lian-Tao Wang}
\author[d]{and Xing Wang}

\affiliation[a]{PITT PACC, Department of Physics and Astronomy, University of Pittsburgh, Pittsburgh, PA 15217, USA}
\affiliation[b]{School of Physics and Astronomy, University of Minnesota, Minneapolis, MN 55455, USA}
\affiliation[c]{Department of Physics, Enrico Fermi Institute, and Kavli Institute for Cosmological Physics, University of Chicago, Chicago, IL 60637, USA}
\affiliation[d]{Department of Physics, University of California at San Diego, La Jolla, CA 92093, USA}

\abstract{
In a previous publication~\cite{Han:2020uak}, 
we showed that a high energy muon collider can make decisive statements about the electroweak (WIMP) Dark Matter (DM), reaching a DM mass which could give the observed thermal relic abundance. In this document, we report new studies of the spin-$0$ minimal WIMP DM at  high energy muon colliders, and update our results on the fermionic spin-$1/2$ case. We find that, by combining multiple inclusive missing mass search channels, it is possible to fully cover the thermal targets of fermionic and scalar doublets, and Dirac triplet, with a  10 TeV muon collider. Higher energies, 14 TeV$-$30 TeV, would be able to cover the thermal targets of Majorana and scalar triplet. For direct discovery of the higher EW multiplets with $n\geq5$, one may need to go beyond a 30 TeV muon collider to fully cover their thermal mass expectation.
}

\makeatletter
\gdef\@fpheader{}
\makeatother

\begin{document}

\maketitle

\noindent
\section{Introduction}

Weakly Interacting Massive Particles (WIMPs), present in many theories beyond the Standard Model (BSM), are natural cold dark matter (DM) candidates \cite{Jungman:1995df,Arcadi:2017kky,Roszkowski:2017nbc}. 
Among the WIMP scenarios, one particularly simple case is the dark matter particle being the lightest member of an electroweak (EW) multiplet. One of the most appealing features of the WIMP scenario is that its mass scale is set by the requirement of saturating the thermal relic abundance, and it is predicted in the range of 1~TeV$-$23~TeV, while lower masses can still be viable with non-thermal production mechanisms. 
The mass splitting among the members of the same multiplet is controlled by the electroweak symmetry breaking, which is small in comparison with the overall mass scale.  
Both the high mass scale and near degeneracy render the DM searches at colliders extremely challenging.
The model-independent mono-$X$ signals ($X=g,\gamma,W/Z, h...$) are not expected to reach a mass beyond two to three hundred GeV at the high luminosity upgrade of the LHC (HL-LHC)~\cite{Low:2014cba,Han:2018wus}.  On the other hand, the disappearing track based searches can extend the coverage up to 900 GeV for a triplet (Wino)~\cite{CidVidal:2018eel}. At the same time, this kind of signature  relies on the mass gap between the members of the EW multiplet. This can introduce additional dependencies, in particular for the case of the Higgsino and the scalar multiplets.
At a future 100 TeV hadron collider, SppC or FCC$_{hh}$, one may hope to extend the coverage to 
a 1.5 (6) TeV for the Higgsino 
(Wino)~\cite{Benedikt:2018csr,CEPCStudyGroup:2018ghi,Strategy:2019vxc}. 

In a previous publication \cite{Han:2020uak}, we showed that a high energy muon collider \cite{muCcoll:2020} can make decisive statements about the electroweak WIMP Dark Matter for a fermionic DM particle in connection with its thermal relic abundance. 
We adopted the benchmark choices of the collider energies and the corresponding integrated luminosities,
\beq
\sqrt{s} = 3,\ 6,\ 10,\ 14, \  30 \ {\rm and}\ 100\ {\rm TeV},\quad {\mathcal L} = 1,\ 4,\ 10,\ 20,\ 90, \ {\rm and}\ 1000\ {\rm ab}^{-1} .
\label{eq:para}
\eeq
We first focus on the universal and inclusive signals, where the particles in an EW multiplet are  produced in association with at least one energetic SM particle. The soft particles or disappearing tracks are treated as invisible. This class of signal is more inclusive and model-independent, since it is independent of the mass splittings within the EW multiplet.The most obvious channel is the pair production of the EW multiplet in association with a photon, which dominates the sensitivity to higher-dimensional EW multiplets.
In addition, we also consider a few other vector boson fusion (VBF) channels unique to a high-energy muon collider \cite{Costantini:2020stv}. 
In particular, the mono-muon channel shows the most promise. 
After considering the inclusive signatures, we also perform a phenomenological estimate of the size of the disappearing track signal.

In this document, we first update the results for the spin-$1/2$ fermionic minimal WIMP DM. We then present our new results for spin-$0$ scalar minimal WIMP DM at high energy muon colliders. 
We find that, by combining multiple missing mass search channels considered here, it is possible to fully cover the thermal targets of fermionic and scalar doublets, and Dirac triplet, with a  10 TeV muon collider. Higher energies, $14-30$ TeV, would be able to cover the thermal targets of Majorana and scalar triplet. For higher multiplets $n\geq5$, one  may need to go beyond a 30 TeV collider to fully cover their thermal mass expectation.
 We reiterate that the high energy muon collider could make a substantial impact in our search for the thermal dark matter, and it should serve as one of the main physics drivers for a high energy muon collider program.

\section{Results}

\subsection{WIMP DM benchmarks}

We consider a broad class of DM candidates, including general SU(2) representations \cite{Cirelli:2005uq,Cirelli:2009uv,DiLuzio:2018jwd}, the so-called ``minimal dark matter'' scenario, for both fermions and scalars. 

More specifically, we will consider multiplets $(1, n, Y)$ under the Standard Model (SM) gauge group SU(3)$_{\rm C} \otimes$SU(2)$_{\rm L} \otimes$U(1)$_{\rm Y}$. 
The $i^{th}$ member of this multiplet has electric charge $Q_i=t^3_i+Y$, where $t^3_i$ is the corresponding SU(2)$_{\rm L}$ isospin component. First, we consider spin-$1/2$ fermionic multiplets. In this case, they only have gauge interactions at the renormalizable level. The mass scale of the EW multiplet is set by the vector-like mass parameter $M$. After electroweak symmetry breaking, the mass spectrum of the multiplet is not exactly degenerate. Minimally, the degeneracy will be lifted by EW loop corrections \cite{Thomas:1998wy,Buckley:2009kv,Cirelli:2005uq,Cirelli:2009uv,Ibe:2012sx}.

\begin{table}[t]
  \centering
    \begin{tabular}{c|c|r|c|c|c|c}
    \hline
    \multicolumn{2}{c|}{\multirow{2}[-3]{*}{Model}} & \multicolumn{1}{c|}{\multirow{2}[-3]{*}{Therm.}} & \multicolumn{4}{c}{2$\sigma$ coverage (TeV)}  \\ \cline{4-7}
    \multicolumn{2}{c|}{$({\rm color}, n, Y)$} & target   & \multicolumn{1}{c|}{mono-$\gamma$} & \multicolumn{1}{c|}{mono-$\mu$} & \multicolumn{1}{c|}{di-$\mu$'s} & \multicolumn{1}{c}{disp. tracks} \\  \hline
    (1,2,1/2)$$ & Dirac & 1.1 TeV & ---  & 1.3  & ---  & $2.3$  \\ \hline  \hline
    (1,3,0)$$ & Majorana & 2.8 TeV & ---   & 1.6  & ---  & $4.8$  \\ \hline
    (1,3,$\epsilon$)$$ & Dirac & 2.0 TeV & 1.2  & 2.0  & 0.6  & $4.8$  \\ \hline  \hline
    (1,5,0)$$ & Majorana & 14 TeV & 4.0  & 2.7  & 2.0  & $4.5$ \\ \hline
    (1,5,$\epsilon$)$$ & Dirac & 6.6 TeV &  4.5 & 2.9 & 2.4  & $4.5$   \\ \hline  \hline
    (1,7,0)$$ & Majorana & 49 TeV  & 4.7   & 3.1   & 3.1  & $4.0$ \\ \hline
    (1,7,$\epsilon$)$$ & Dirac & 16 TeV & 4.8  & 3.2 & 3.4 & $4.1$  \\ \hline  \hline
    \end{tabular}%
    \caption{
    Minimal fermionic dark matter candidates considered in this paper and a brief summary of their $2\sigma$ coverage at a 10 TeV high energy muon collider with the three individual channels. Further details of individual and combined channels, the $2\sigma$ reaches, and different collider parameter choices,  including $\sqrt s=$3, 6, 10, 14, 30, 100~TeV are provided in the summary plots in \autoref{fig:summaryF}.
    \label{tab:WIMP} 
    }
\end{table}%

Both real and complex scalar EW multiplets can contain viable dark matter candidates. The discussion of EW loop corrections to the mass splitting parallels to that of the fermions. 
One main difference is that the scalar can have more couplings in addition to gauge interactions at the renormalizable level, of the form $\chi \chi^\dagger H H^{\dagger}$ with different ways of contracting SU(2)$_{\rm L}$ indices. Such couplings can induce sizable splittings in the EW multiplet after the EW symmetry breaking, and also introduce extra contributions to the annihilation cross sections. Hence, there are more parameters and model dependences in comparison with the case of fermionic EW multiplets, and we assume them to be negligible.

\begin{table}[t]
  \centering
    \begin{tabular}{c|c|r|c|c|c|c}
    \hline
    \multicolumn{2}{c|}{\multirow{2}[-3]{*}{Model}} & \multicolumn{1}{c|}{\multirow{2}[-3]{*}{Therm.}} & \multicolumn{4}{c}{2$\sigma$ coverage (TeV)}  \\ \cline{4-7}
    \multicolumn{2}{c|}{$({\rm color}, n, Y)$} & target   & \multicolumn{1}{c|}{mono-$\gamma$} & \multicolumn{1}{c|}{mono-$\mu$} & \multicolumn{1}{c|}{di-$\mu$'s} & \multicolumn{1}{c}{disp. tracks} \\  \hline
    (1,2,1/2)$$ & Complex & 0.54 TeV & ---  & 0.7  & ---  & $1.8$  \\ \hline  \hline
    (1,3,0)$$ & Real & 2.5 TeV & ---   & 0.9  & ---  & $2.1$  \\ \hline
    (1,3,$\epsilon$)$$ & Complex & 1.6 TeV & ---  & 1.2  & ---  & $2.3$  \\ \hline  \hline
    (1,5,0)$$ & Real & 15 TeV & 1.2  & 2.0  & 0.9  & $2.7$ \\ \hline
    (1,5,$\epsilon$)$$ & Complex & 6.6 TeV &  2.2 & 2.3 & 1.3  & $2.9$   \\ \hline  \hline
    (1,7,0)$$ & Real & 54 TeV  & 3.1   & 2.6   & 2.0  & $3.1$ \\ \hline
    (1,7,$\epsilon$)$$ & Complex & 16 TeV & 3.6  & 2.8 & 2.4 & $3.2$  \\ \hline  \hline
    \end{tabular}%
    \caption{Minimal scalar dark matter candidats, their thermal target masses, and a brief summary of their $2\sigma$ coverage at a 10 TeV high energy muon collider with the three individual channels. 
    Further details of individual and combined channels, the $2\sigma$ reaches, and different collider parameter choices,  including $\sqrt s=$3, 6, 10, 14, 30, 100~TeV are provided in the summary plots  \autoref{fig:summaryS}.
       \label{tab:WIMPS}
       }
\end{table}

The mass of the EW multiplet and its interactions with the SM particles determine the thermal relic abundance of the cold DM. In the minimal scenarios considered in this paper, the EM multiplets only have  SM gauge interactions. Hence, requiring thermal relic abundance matches the observation \cite{Aghanim:2018eyx} can determine the mass of the dark matter, which we refer to as the {\it thermal target}. 
We list the multiplets according to the SM gauge quantum numbers under SU(3)$_{\rm C}\otimes$SU(2)$_{\rm L} \otimes$U(1)$_{\rm Y}$ and the predicted thermal targets 
in \autoref{tab:WIMP} for fermionic WIMP DM and in \autoref{tab:WIMPS} for scalar WIMP DM, 
which set the benchmark for searches at future colliders. We note here that perturbative calculation of the thermal targets for many of the EW multiplets receives large corrections from the Sommerfeld enhancement~\cite{Belotsky:2005dk,Hisano:2006nn,Cirelli:2007xd} as well as bound state effects~\cite{An:2016gad,Mitridate:2017izz}. 
In detail, we follow Ref.~\cite{Bottaro:2021snn} in thermal target calculation for real representations for scalar and fermion, where both sommerfeld and bound-state corrections have been taken into account. The thermal masses for the complex represetations are taken from Ref.~\cite{DelNobile:2015bqo}, where only sommerfeld corrections are partially included. 
We note that these thermal targets have some theoretical uncertainties due to the non-perturbative effects mentioned above. Nevertheless, they can serve as useful targets. 
Reaching these targets marks a great triumph for future colliders in probing WIMP dark matter, with the potential of the next milestone discovery. 

\subsection{Search channels}

With only gauge interactions,
the production and decay for the EW multiplets are highly predictable. Since the mass splittings  between the charged and neutral states are expected to be small, typically of the order of a few hundred MeV from EW loops in the minimal scenario, the decay products will be very soft, most likely escaping the detection. In this case, the main signal at high energy muon colliders is large missing energy-momenta. We note that, unlike in the high energy hadronic collisions where only the missing transverse momenta can be reconstructed by the momentum conservation, the four-momentum of the missing particle system can be fully determined in leptonic collisions because of the well-constrained kinematics. Importantly, a large missing invariant mass can be inferred. We thus introduce the ``missing mass'' defined as
\begin{equation}
m^2_{\rm missing} \equiv (p_{\mu^+} + p_{\mu^-} - \sum_i p_i^{\rm obs})^2,
\label{eq:recoilM}
\end{equation}
where $p_{\mu^+}, p_{\mu^-}$ are the momenta for the initial colliding beams, and $p_i^{\rm obs}$ is the momentum for the $i^{th}$ final state particle observed. If the EW multiplet particles are not detected, $m_{\rm missing}$ for the signal will have a threshold at twice the dark matter mass. We thus call this characteristic signature the ``missing-mass'' signal. 

\subsubsection{Mono-Photon}
\label{sec:monophoton}

We first consider the mono-photon signal. The  members of the electroweak multiplet, both charged and neutral,  can be produced either via $s$-channel $\gamma$ and $Z$ or via the vector boson fusion processes. 
We consider the following signal processes
\begin{eqnarray}
 \mu^+ \mu^-  &\rightarrow& \gamma \chi \chi \quad {\rm via\ annihilation}\ \mu^+\mu^- \to \chi\chi , \\
\gamma \gamma & \rightarrow& \gamma \chi \chi \quad {\rm via}\ \gamma\gamma\to \chi\chi , \\
 \gamma \mu^\pm & \rightarrow &\gamma \nu \chi \chi \quad {\rm via}\ \gamma W\to \chi \chi, \\
 \mu^+ \mu^- & \rightarrow& \gamma \nu \nu \chi \chi \quad {\rm via}\ WW\to \chi\chi\ {\rm and}\ 
 \mu^+\mu^- \to \chi\chi Z.
\end{eqnarray}
where $\chi$ represents any state within the $n$-plet and $\chi\chi$ represents any combination of a pair of the $\chi$ states allowed by the gauge symmetries.

As for the signal identification, we first require a photon in the final state within the detector acceptance 
\begin{equation}
10^\circ < \theta_\gamma < 170^\circ .
\label{eq:angle}
\end{equation}
Taking into account the invariant mass of the dark matter pair system being greater than $2 m_\chi$, we impose further selection cuts on the energy of the photon and on the missing mass 
\begin{equation}
 E_\gamma > 25~{\rm GeV} ,\ \ \ 
m^2_{\rm missing} \equiv (p_{\mu^+} + p_{\mu^-} - p_\gamma)^2 > 4m_\chi^2 .
\label{eq:ecut}
\end{equation}
The missing-mass cut is equivalent to an upper limit on the energy of the photon $E_\gamma < (s - 4 m_\chi^2)/2\sqrt{s}$, 
where $\sqrt{s}$ is the collider c.m.~energy.
We consider multiple sources of the SM background and the most significant SM background, after the selection cuts, is
\begin{equation}
\mu^+ \mu^- \rightarrow \gamma \nu \bar{\nu}, 
\end{equation}
dominantly from contributions via the $t$-channel $W$-exchange.

In the reach projection, we take a conservative approach to  estimate the significance as 
\bea
{\rm N_{\rm SD}} = {S\over \sqrt{S+B + (\epsilon_S S)^2 + (\epsilon_B B)^2}},
\label{eq:NSD}
\eea
where $S$ and $B$ are the numbers of events for the signal and background, and $\epsilon_S$ and $\epsilon_B$ are the corresponding coefficients for systematic uncertainties, respectively. It is clear from this equation that, in a statistical uncertainty-dominated scenario ($\epsilon_S=\epsilon_B=0$), the significance scales as $S/\sqrt{S+B}$, and in a systemic uncertainty-dominated scenario, the significance scales as $S/(\epsilon_B B)$. In processes where the $S/\sqrt{B}$ is high, but $S/B$ is tiny, one needs to pay special attention to the uncertainty arising from the systematics.

\subsubsection{Mono-Muon}
\label{sec:monomuon}
While the mono-photon is a generic dark matter signal for all high energy colliders, mono-muon signal to be studied in this section is unique to muon colliders. 
The leading signal processes are 
\begin{equation}
\aligned
\gamma \;\mu^\pm & \rightarrow \mu^\pm \chi \chi\quad {\rm via}\ \ \gamma Z \to \chi \chi ,\\ 
\mu^+ \mu^- &\rightarrow \mu^\pm \nu \chi \chi \quad {\rm via}\ \ \gamma W, ZW \to \chi \chi,
\endaligned
\end{equation}
where $\chi$'s represent any states within the $n$-plet, and $\chi\chi$ represents any combination of a pair of the $\chi$ states allowed by gauge symmetries. The $\mu^\pm$ is required to be in the detector coverage as in \autoref{eq:angle}.

The main background comes from processes in which a charged particle (mostly muon) escapes detection in the forward direction, due to the finite angular acceptance of the detector. The dominant process is 
\begin{equation}
\gamma \;\mu^\pm \rightarrow \mu^\pm \nu \bar{\nu},
\label{eq:gmutomununu}
\end{equation}
resulting from both $Z\to \nu \bar\nu$ and $W\to \mu\bar\nu$, where the muon from which the photon radiates missed the detection.

We will impose the missing mass cut
\begin{equation}
m^2_{\rm missing} = (p^{\rm in}_{\mu^+} + p^{\rm in}_{\mu^-} - p^{\rm out}_{\mu^\pm})^2 > 4m_\chi^2.
\end{equation}
and
\begin{equation}
E_{\mu^\pm} > 0.71,\; 1.4,\; 2.3,\; 3.2,\; 6.9,\; 22.6~{\rm TeV},\quad {\rm for }~\sqrt{s}=3,\; 6,\; 10,\; 14,\; 30,\; 100~{\rm TeV},
\end{equation}
to suppress the background

\subsubsection{VBF Di-Muon}
\label{sec:VBF}
Beyond the single muon signature, one could also consider to tag both muons in the final state to account for other additional contributing channels via the VBF
\begin{equation}
\mu^+ \mu^- \rightarrow \mu^+ \mu^- \chi \chi \quad {\rm via\ fusion}\ \ \gamma^*\gamma^*, \gamma^* Z, ZZ \to \chi \chi 
\end{equation}
where $\chi$ represents any state within the $n$-plet.
We require both final state muons to be in the detector coverage as in \autoref{eq:angle}. This effectively suppresses the backgrounds that are dominated by low momentum transfer.  For a $\gamma^*$ initiated process, the cross section with finite angle scattering falls at higher energies of the final state muons as $1/(p_T^\mu)^2$ for each tagged muon.
Although $p_T^{\rm muon} \sim M_Z$ for a $Z$-initiated process, the muons can still be highly boosted due to the large beam energy, with a scattering angle of the order 
$\theta_\mu \sim M_Z/E_{\mu f}$~\cite{Han:2020pif}, likely outside the detector coverage.

The leading irreducible background is
\begin{equation}
\mu^+ \mu^- \rightarrow \mu^+ \mu^- \nu \bar{\nu}.
\end{equation}
The dominant contributions are from both $\gamma^*\gamma^*, \gamma^* Z, ZZ$ fusion processes as well as $ZZ\to \mu^+ \mu^-\ \nu\bar \nu$.
 To suppress the large non-fusion background primarily from a $Z$ decay to leptons,  the muons are required to have
\begin{equation}
m_{\mu^+\mu^-} > 300~{\rm GeV} ,\quad 
m_{\rm missing} = (p^{\rm in}_{\mu^+} + p^{\rm in}_{\mu^-} - p^{\rm out}_{\mu^+} - p^{\rm out}_{\mu^-})^2 > 4m_\chi^2 .
\end{equation}

\subsubsection{Disappearing tracks and other signatures}
\label{sec:DT}
Besides the conventional missing mass search, the loop-induced mass splitting among the component states of the EW multiplet also results in a disappearing track signature which can be used to enhance the reach.

The small mass splitting and anticipated lifetimes allow us to develop the following very {\it simple} strategy for a phenomenological estimation for the signal rate.
First, the charge $\pm1$ states will have macroscopic lifetime from the collider perspectives, generating the signature of ``disappearing tracks'' typically associated with long-lived particles. Second, although the doubly charged state in the $Y=0$ multiplets has a lifetime as large as $0.5$~mm, it would be difficult to reach the tracker due to the typical low boost of $\gamma=E_\chi/m_\chi$ for a heavy $\chi$ at a muon collider.\footnote{We discuss the potential double displacement signature in the last part of this section.}As a result,  the decay of  states with a charge $\pm 2$ or more into the lower charged states can be treated as  {\it prompt}, and only the charge $\pm 1$ states have a relevant long lifetime. Hence, all the EW pair productions considered in the previous sections, including the production of the states with charge $\geq 2$,  gives rise to  long-lived charged $\pm 1$ particles in the final state, with typical proper decay lengths
\begin{equation}
  c\tau =
    \begin{cases}
      0.64~{\rm cm} & \text{EW doublet}\\
      \frac{8}{(n-1)(n+1)}\times5.7~{\rm cm} & \text{EW odd $n$-plet}
    \end{cases}       
\end{equation}
We emphasize that these lifetimes are based on mass splittings generated by one loop electroweak corrections. In principle, in more extended models, there could be additional corrections even if the DM multiplet remains the lightest. For example, the Higgsino multiplet can receive additional contributions from dimension five operators. In the case of the scalar dark matter, there are renormalizable couplings of the form $\chi \chi^\dagger H H^\dagger $ which can give rise to even larger contribition. All of these can change the disappearing track signal in a significant way. Hence, in our discussion of disappearing track signal, we should keep in mind we are focusing on the minimal model by assumption. At the same time, the inclusive missing mass signal discussed earlier can be applicable even in non-minimal cases as long as the mass splittings are not very large.

The disappearing track signature can be reconstructed in collider experiments via a series of inner tracker hits that are not followed by hits in the outer layers, which can form a track with a consistent curvature. We assume the reconstruction probability of a signal event with one disappearing track is
\beq
\epsilon_\chi (\cos\theta,\gamma, d_T^{\rm min})=\exp\left(\frac {-d_T^{\rm min}} {\beta_T \gamma c\tau}\right),\quad \text{with}~d^{\rm min}_T=5\ {\rm cm}~{\rm and~} |\eta_\chi|<1.5
\label{eq:dtcut}
\eeq
where $\gamma=E_\chi/m_{\chi}$
and 
$\beta_T=\sqrt{1-1/\gamma^2} \sin\theta$, which is the transverse velocity in the lab frame. 
The minimal transverse displacement of $5$~cm represents the minimal track reconstruction requirement (of two hits) for a typical muon collider detector design with pixel layers.\footnote{For instance, in Ref.~\cite{ILC:2007vrf,Capdevilla:2021fmj}, the detetor layout has the two inner most pixel layers with radii 3.1~cm and 5.1~cm.} Future detector design studies could further optimize the layout for Higgsino-like short tracks, e.g., moving pixel layers closer to the beam spot, that could greatly improve the the muon collider sensitivities.

A unique challenge for a muon collider in identifying the disappearing track signal is the high level of the beam-induced background (BIB). Preliminary studies~\cite{DiBenedetto:2018cpy,Bartosik:2019dzq,Bartosik:2020xwr} (based on a 1.5 TeV muon collider) have demonstrated that more than ${\mathcal O} (10^2)$ hits per cm$^2$ are expected at the first layer. Detector simulation for a 10 TeV muon collider also suggests ${\mathcal O} (10^2)$ background events \cite{Capdevilla:2021fmj}. Here, we use {\it 20 (50)} identified signal events for $2 \sigma~(5 \sigma)$ reach, which would be consistent with around 100 background events, as performance benchmarks for exclusion and discovery for the mono-photon plus disappearing track searches. We require these amount of signal events after imposing the mono-photon selection cuts (\autoref{eq:angle} and \autoref{eq:ecut}) and the disappearing track selection cuts (\autoref{eq:dtcut}).

\begin{figure}[tb]
\centering
\includegraphics[width = 0.7\textwidth]{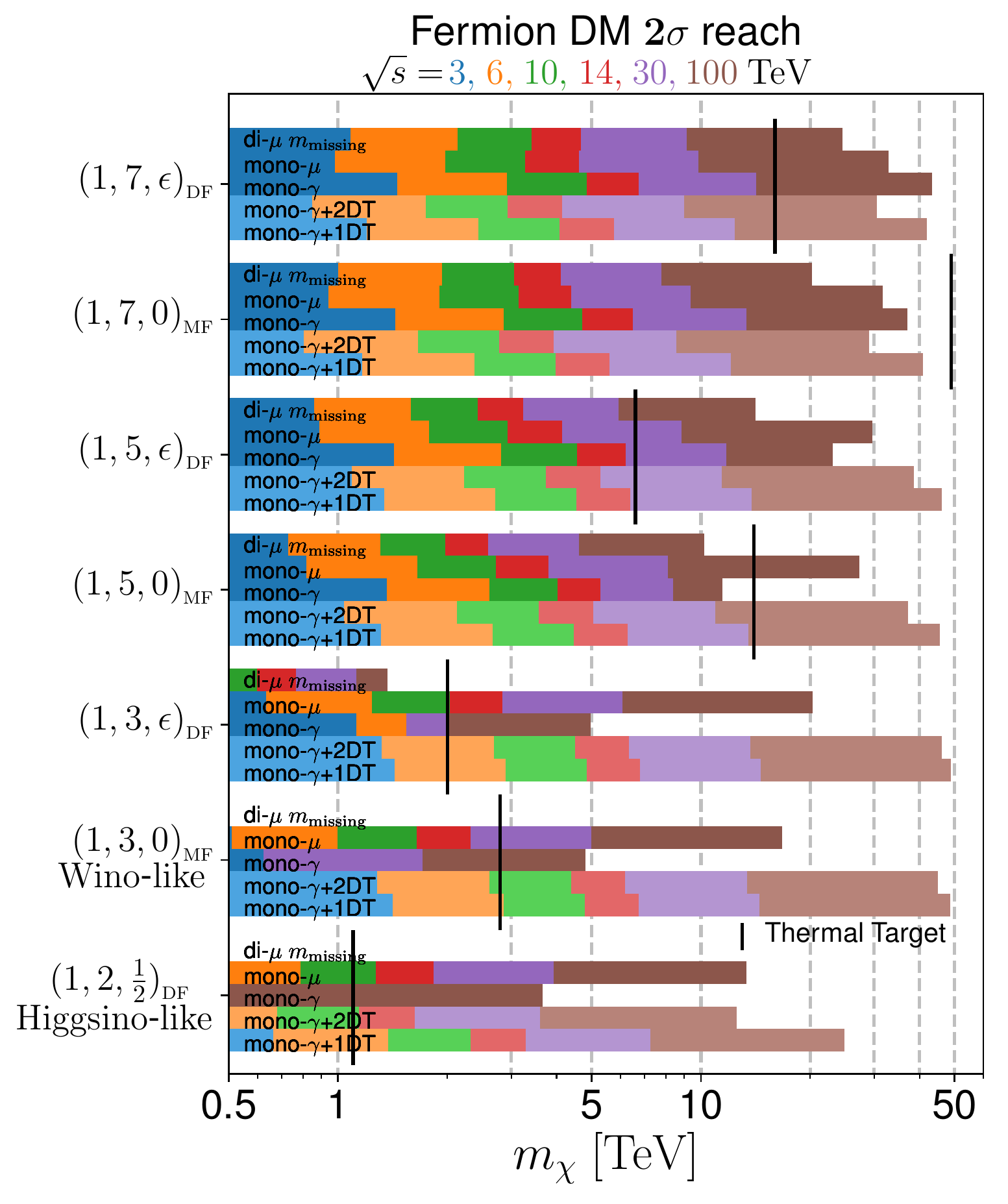}
\caption{$2\sigma$ exclusion of fermion DM masses with horizontal bars for individual search channels and various muon collider running scenarios by the different color codes. 
The vertical bars indicate the thermal mass targets for the corresponding WIMP DM. }
\label{fig:summaryF}
\end{figure}


\begin{figure}[tb]
\centering
\includegraphics[width = 0.7\textwidth]{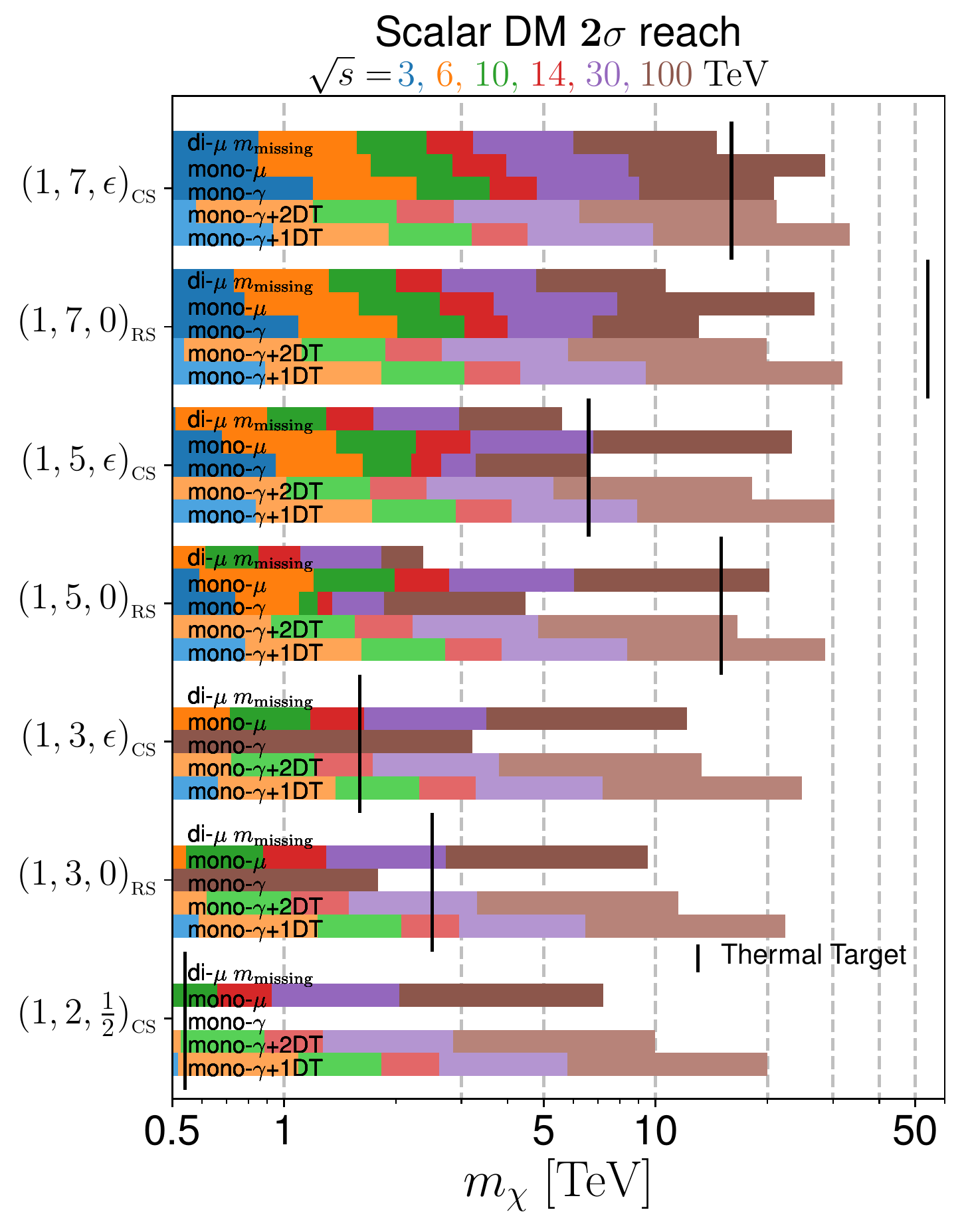}
\caption{Same as Fig.~1 for scalar DM.
}
\label{fig:summaryS}
\end{figure}


\subsection{Projected sensitivity}

We present the sensitivity results for the WIMP DM of Dirac fermion (DF), Majorana fermion (MF), complex scalar (CS), and real scalar (RS) in  Fig.~\ref{fig:summaryF} and~\ref{fig:summaryS}. The reaches for $2\sigma$ exclusion are shown for individual search channels, and various muon collider running scenarios are indicated by the color codes:
\begin{itemize}
    \item 
    The mono-muon channel, a unique signal for muon collider, shows a lot of potential, and is especially promising for lower-dimensional EW multiplets with $n\leq 3$.
    \item
    The traditional mono-photon channel at lepton colliders is suitable for higher-dimensional EW multiplets, due to the coupling enhancement for high EW $n$-plets and  the high multiplicity of the final state. In principle, one can consider radiation of other EW gauge bosons such as $W$ and $Z$ to further improve the sensitivity~\cite{Bottaro:2021snn,IMCCphysicsreport}.
    \item
    Disappearing track will play an indispensable role in the search for EW multiplets. The mono-photon channel with one disappearing track will have the largest signal rate and can extend the reach significantly for all odd-dimensional cases. Requiring disappearing-track pairs will reduce the reach. However, it is a cleaner signal and could be more important if the single disappearing track signature does not provide enough background suppression.
\end{itemize}

\begin{figure}[tb]
\centering
\includegraphics[align=c, width = 0.7\textwidth]{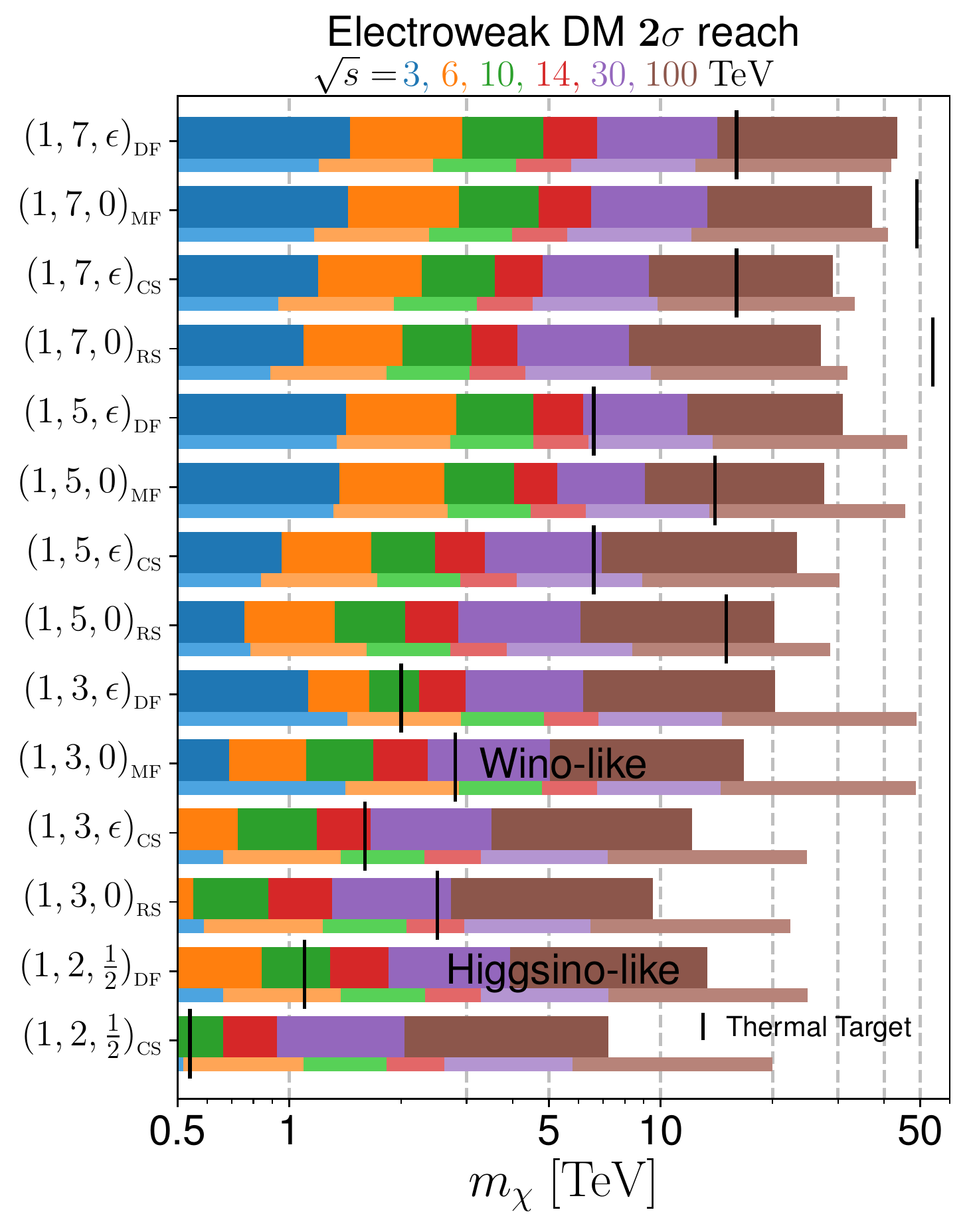}
\caption{$2\sigma$ exclusion of DM masses with horizontal (thick) bars for combined channels and various muon collider running scenarios by the different color codes. 
The thin bars are the estimation of the mono-photon plus one disappearing track search. 
The vertical bars indicate the thermal mass targets for the corresponding WIMP DM. 
}
\label{fig:summaryAll}
\end{figure}

\begin{figure}[tb]
\centering
\includegraphics[align=c, width = 0.7\textwidth]{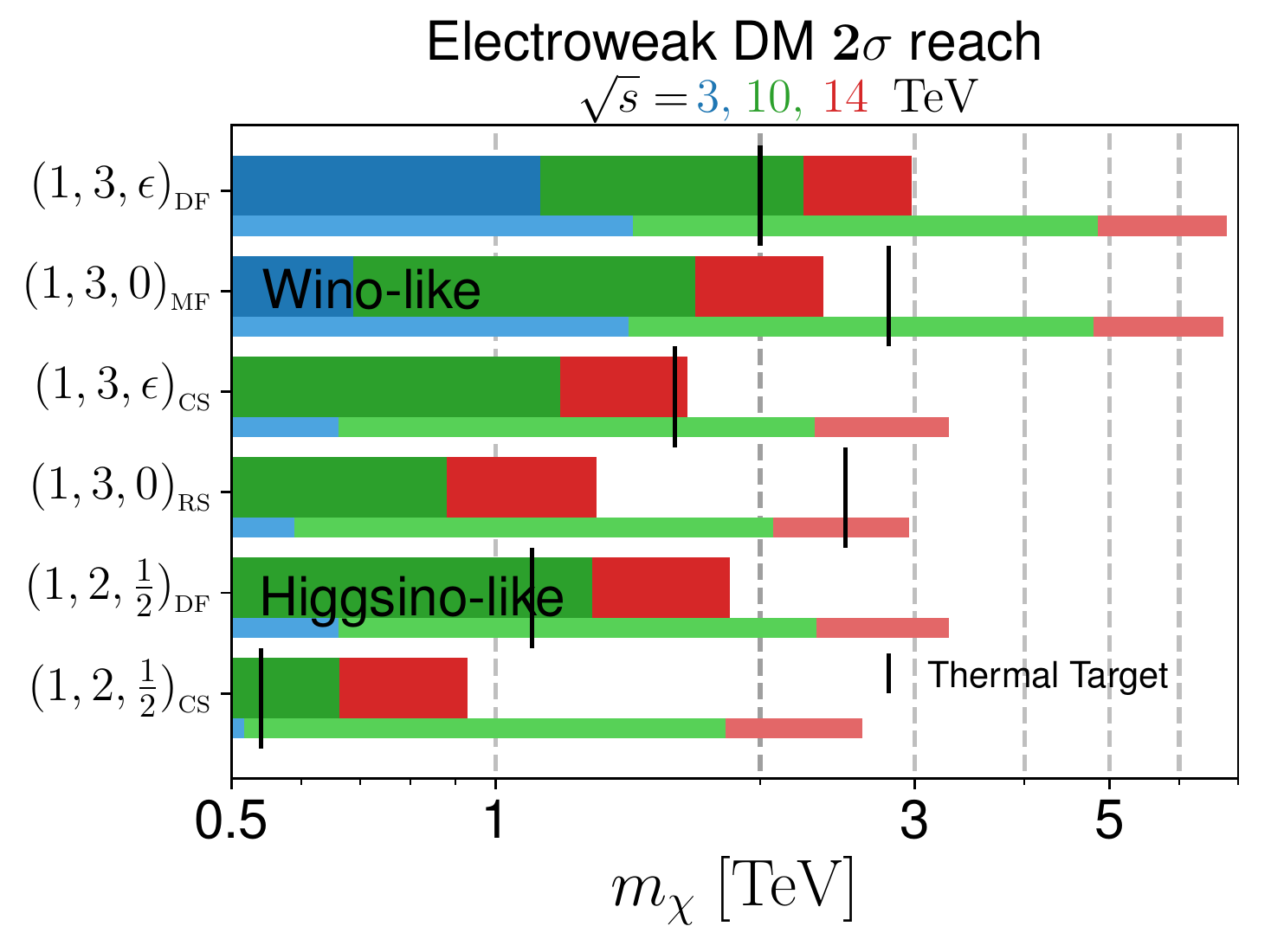}
\caption{Same as Fig.~\ref{fig:summaryAll}, a zoom-in version for $\sqrt s = 3, 10$ and 14 TeV.}
\label{fig:summaryAll2}
\end{figure}

The $2\sigma$ reaches for fermionic and scalar DM are summarized in Fig.~\ref{fig:summaryAll}. A zoom-in  version with fewer energies of $\sqrt s=3, 10$ and 14 TeV is provided in Fig.~\ref{fig:summaryAll2} for reader's convenience.
The thick (darker) bars represent the reach in DM mass (horizontal axis) by combining different inclusive missing-mass signals. The thin (fainter) bars are our estimates of the mono-photon plus one disappearing track search. 
For comparison, we have also included the target masses (vertical bars in black) for which the dark matter thermal relic abundance is saturated by the EW multiplets DM  under consideration.  When combining the inclusive (missing mass) channels, the overall reach is less than the kinematical limit $m_\chi \sim \sqrt{s}/2$, especially for EW multiplets with $n \leq 3$ due to the low signal-to-background ratio. It is possible to cover (with $2 \sigma$) the thermal targets of the doublet and Dirac fermion triplet with a 10 TeV muon collider. The complex scalar tripet can be covered by a 14 TeV muon collider.  For the real scalar and Majorana fermion triplet, a 30 TeV option would suffice. 
The thermal targets of complex scalar and Dirac fermion (real scalar and Majorana fermion) 5-plet would be covered by 30 (100) 
TeV muon colliders. The 100 TeV option will also cover the thermal target for the complex scalar and Dirac fermion 7-plet. The real scalar and Majorana fermion 7-plet can be probed up to $30-40$ TeV in mass at a 100 TeV muon collider, with their thermal target still out of reach. It is important to emphasize that, in order to cover the thermal target, the necessary center of mass energy and luminosity in many cases can be much lower than the benchmark values we showed in \autoref{eq:para}. 
At the same time, the disappearing track signal has excellent potential, and could be the leading probe for 5-plet or lower EW multiplet. Based on our study, it could bring the reach very close to the kinematical threshold $m_\chi \sim \sqrt{s}/2$. We note here, a 6 TeV muon collider with disappearing track search can cover the thermal target of the doublet case, motivating further detailed studies in this direction. 
In principle, a 3 TeV muon collider has sufficient energy to kinematically access the pure-Higgsino DM through the disappearing track channel. However, with the current detector layout design~\cite{ILC:2007vrf} and the short lifetime, the signal efficiency would still be too low~\cite{Han:2020uak}. 
The maximal signal efficiency can be estimated as follows. At $E_{\rm CM} = 3 $ TeV, the Higgsino would be produced quite close to the threshold. With a lifetime of 0.02~ns, it would have a lab frame lifetime smaller than 0.56~cm, with a smaller {\it transverse} displacement. The single disappearing track reconstruction would have an efficiency at most $2.5\times 10^{-4}$ without taking into account any experimental acceptance. The Higgsino production rate without the requirement of the existence of a 25~GeV $p_T$ photon is 10~fb. After requiring such photon in association with the single track, the cross section is 1~fb. Higgsino will be produced with a distribution of pseudo rapidity, yielding an even smaller number of signal events. All these points towards less than 1 signal event. At the same time, the background would be around 20 from 
BIB.\footnote{Our findings disagree with the results in Ref.~\cite{IMCCphysicsreport}, based upon Ref.~\cite{Capdevilla:2021fmj}. After communicating with the authors of those papers, they have informed us of a numerical error in their Higgsino analysis, and we look forward to further comparative studies. 
}
The results are also summarized in Tables \ref{tab:WIMP} and \ref{tab:WIMPS} for a 10 TeV muon collider.
There are a few considerations that could potentially improve the detection sensitivity. We could increase the luminosity by a factor of 8 (see Fig.17(a) of Ref.~\cite{Han:2020uak} to enable exclusion from the mono-photon channel), or design the  detector with pixel layers closer to the beam spot, and/or increase the center of mass energy to provide more boost that increases lifetime in the lab-frame.

\section{Executive Summary}
\label{sec:Sum}

In this document, we reported our new studies of the spin-$0$ minimal WIMP DM at  high energy muon colliders, and updated our previous results on the fermionic spin-$1/2$ case. 
By combining multiple inclusive missing mass search channels, it is possible to fully cover the thermal targets of fermionic and scalar doublets, and Dirac triplet, with a  10 TeV muon collider. Higher energies, 14 TeV$-$30 TeV, would be able to cover the thermal targets of Majorana and scalar triplet. For direct discovery of the higher EW multiplets with $n\geq5$, one may need to go beyond a 30 TeV muon collider to fully cover their thermal mass expectation. Our studies on the disappearing tracks, as a first look, identify cases and regions where potential gains in reach can be significant. 

Overall, our results have demonstrated that muon colliders running at multi-TeV energies have great potential in searching for the EW multiplets and can make a decisive statement about their viability as WIMP dark matter candidates. This should serve as a main physics driver for the high energy muon colliders.

\begin{acknowledgments}
The work of TH was supported in part by the U.S.~Department of Energy under grant 
No.~DE-SC0007914 
and in part by the PITT PACC. 
ZL was supported in part by the U.S. Department of Energy (DOE) under grant No. DE-SC0022345.
LTW was supported by the DOE grant de-sc0013642. 
XW was supported by the National Science Foundation under Grant No.~PHY-1915147.
\end{acknowledgments}

\clearpage

\bibliography{refs_MuCDM}

\bibliographystyle{utphys}

\end{document}